\documentclass{elsart4}
\catcode`\@=11
\def\@jou@vol@pag{\small to appear in JMMM (proceedings of JEMS'04) \quad
 version of 30 June 2004}
\let\@j@v@p\@jou@vol@pag
\let\@@j@v@p\@jou@vol@pag
\catcode`@=12

 \usepackage{graphicx}

\usepackage{amssymb}

\begin{document}

\def\Vec#1{\mathbf{#1}}

\def\GS{GS}

\def\Section#1{}
\def\etal#1{ {\it et al.}}

\begin{frontmatter}

\title{Exchange constants and spin dynamics in Mn$_{12}$-acetate}

\author[BS]{A. Honecker}
\author[BS]{N. Fukushima\thanksref{NFpres}}
\thanks[NFpres]{Present address:
University of the Saarland, Theoretische Physik, Geb.\ 38,
66123 Saarbr\"ucken, Germany}
\author[Fri]{B. Normand}
\author[CEA]{G. Chaboussant}
\author[Bern]{H.-U. G\"udel}
\address[BS]{Institut f\"ur Theoretische Physik,
          Technische Universit\"at Braunschweig,
          Mendelssohnstr.\ 3, 38106 Braunschweig, Germany}
\address[Fri]{D\'epartement de Physique, Universit\'e de Fribourg,
          1700 Fribourg, Switzerland}
\address[CEA]{Laboratoire L\'eon Brillouin, CEA-Saclay,
           91191 Gif-sur-Yvette Cedex, France}
\address[Bern]{Departement f\"{u}r Chemie und Biochemie,
          Universit\"{a}t Bern, Freiestrasse 3, 3000 Bern 9, Switzerland}

\begin{abstract}

We have obtained new inelastic neutron scattering (INS) data for the molecular
magnet Mn$_{12}$-acetate which exhibit at least six magnetic peaks in the
energy range 5--35~meV. These are compared with a microscopic
Heisenberg model for the 12 quantum spins localised on the Mn ions,
coupled by four inequivalent magnetic exchange constants.
A fit to the magnetic susceptibility under
the constraint that the spin of the ground state be $S=10$ yields
two dominant exchange constants of very similar value,
$J_1 \approx J_2 \approx 65$~K ($\approx 5.5$~meV),
and two smaller exchange constants $J_3$ and $J_4$.
We compute the low-lying excitations
by exact numerical diagonalisation
and demonstrate that the parameters determined from the
ground state and susceptibility fit provide qualitative
agreement with the excitations observed by INS.

\end{abstract}

\end{frontmatter}

\Section{Introduction}

Magnetic molecules present a fascinating new class of materials with a
wide variety of applications (for a recent review see \cite{Schnack}).
Coherent quantum phenomena in these mesoscopic systems are one focus
of recent research \cite{TuBa02}. Despite being among the first generation
of molecular magnets to be synthesised, Mn$_{12}$-acetate \cite{Sessoli93}
remains that with the largest barrier to thermally activated tunnelling.
Although much work has been devoted to Mn$_{12}$-acetate over the past
decade, the microscopic mechanisms for the observed low-energy phenomena
have remained controversial.

\begin{figure}[ht]
\includegraphics[width=\columnwidth]{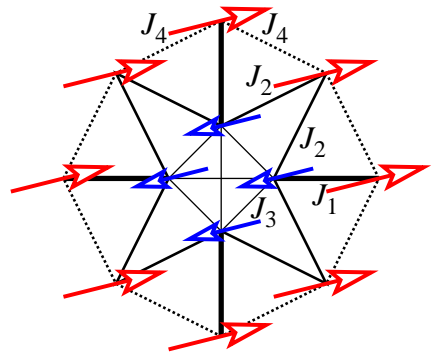}
\caption{Magnetic exchange model for Mn$_{12}$-acetate.
Arrows denote the twelve Mn ions: eight Mn$^{3+}$ ions on the crown have
local spin $S=2$ while four Mn$^{4+}$ ions in the core have $S = 3/2$.
Lines show exchange paths with interaction parameters
$J_1$, $J_2$, $J_3$ and $J_4$.}
\label{fig_model}
\end{figure}

Here we discuss a microscopic exchange model for Mn$_{12}$-acetate.
Twelve quantum spins $\Vec{S}_i$ are coupled by Heisenberg exchange
interactions
\begin{equation}
H = \sum_{i,j} \, J_{ij} \, \Vec{S}_i \cdot \Vec{S}_j
\label{eqMod}
\end{equation}
with four different exchange constants $J_1$, $J_2$, $J_3$ and $J_4$,
as represented in Fig.\ \ref{fig_model}.

Many experimental studies, including inelastic neutron scattering (INS)
\cite{Mirebeau99}, show that the ground state (\GS) of Mn$_{12}$-acetate
has total spin $S = 10$. This may be rationalised by considering
an arrangement of 8 parallel spins $S = 2$ on the crown
Mn$^{3+}$ ions oriented antiparallel to 4 aligned $S = 3/2$ spins
on the core Mn$^{4+}$ ions (Fig.\ \ref{fig_model}).
The $S = 10$ \GS\ imposes a strong constraint on the allowed
exchange constants in (\ref{eqMod}), excluding \cite{Raghu01,we04}
a number of parameter sets proposed in the literature, such as that
obtained by the {\em ab initio} local density approximation
\cite{Boukhvalov02}.

\Section{Magnetic susceptibility}

\begin{figure}[ht]
\centering
\includegraphics[width=\columnwidth]{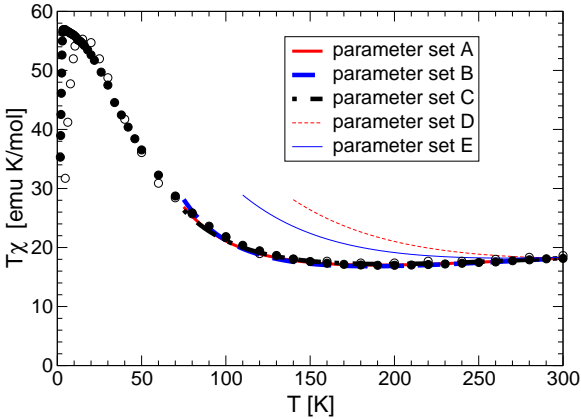}
\caption{Static magnetic susceptibility.
Filled and open circles are measured respectively on a deutered sample
under a field of 0.1~T \cite{we04} and on a non-deuterated sample at
1~T~\cite{Schake94}. Lines are obtained from an 8th-order high-temperature
series for the parameter sets in Table \ref{TabSpec}.}
\label{fig:susc}
\end{figure}

The magnetic susceptibility $\chi$ is a valuable quantity in the
determination of magnetic exchange constants. Fig.\ \ref{fig:susc}
shows two results for $\chi$, measured with an ordinary sample under
an applied field of 1~T \cite{Schake94}, and with a deuterated sample at
0.1~T \cite{we04}. Both data sets agree well for temperatures between
40 and 300~K despite the different conditions, demonstrating the
reliability of the susceptibility measurement at high temperatures.
Exchange constants can then be determined by comparison with
a symbolic high-temperature series expansion. In combination
with a numerical test of the $S = 10$ ground-state requirement,
this restricts the possible parameters to a narrow region around
$J_1 \approx J_2 \approx 60$~K, $J_3 \approx J_4 = 5$--10~K \cite{we04}.

Columns A, B and C of Table \ref{TabSpec} list three choices of parameters
in this region (parameter set A was used in \cite{we04}). Columns
D and E contain the parameter sets proposed in \cite{Regnault02} and
\cite{Park04} respectively. The lines in Fig.~\ref{fig:susc} show
the susceptibility $\chi$ obtained from an average of four different
Pad\'e approximants to the 8th-order high-temperature series \cite{we04}
evaluated with the corresponding parameters. The last row of Table
\ref{TabSpec} lists the effective $g$-factor entering the absolute
value of $\chi$ (electron paramagnetic resonance \cite{Barra97} yields
$g_{\rm eff} = 1.968$). Parameter sets A, B and C yield good agreement
with the experimental results, whereas the results for sets D and E
are in clear disagreement. We conclude that the exchange constants
proposed in Refs.~\cite{Regnault02,Park04} are incompatible with $\chi$.

\begin{table*}[htb]
\caption{Energy $E$ and symmetry $k$ of low-lying excitations for a
Mn$_{12}$-acetate exchange model with different parameter sets. Exchange
constants are given in Kelvin [K]. The \GS\ has spin $S=10$ in all five cases.
No energetic correction is applied for uniaxial anisotropy of the cluster. }
\label{TabSpec}
\centering
\begin{tabular}{l|cc|cc|cc|cc|cc}
\hline
& \multicolumn{2}{c|}{(A) \cite{we04}}
 & \multicolumn{2}{c|}{(B)}
  & \multicolumn{2}{c|}{(C)}
   & \multicolumn{2}{c|}{(D) \cite{Regnault02}}
    & \multicolumn{2}{c}{(E) \cite{Park04}} \\
& \multicolumn{2}{c|}{$J_1 = 67.2$, $J_2 = 61.8$,}
  & \multicolumn{2}{c|}{$J_1 = 64.5$, $J_2 = 60.3$,}
    & \multicolumn{2}{c|}{$J_1 = 64$, $J_2 = 65$,}
      & \multicolumn{2}{c|}{$J_1 = 119$, $J_2 = 118$,}
        & \multicolumn{2}{c}{$J_1 = 115$, $J_2 = 84$,} \\
& \multicolumn{2}{c|}{$J_3 = 7.8$, $J_4 = 5.6$}
  & \multicolumn{2}{c|}{$J_3 = 4.2$, $J_4 = 6.3$}
    & \multicolumn{2}{c|}{$J_3 = 11$, $J_4 = 4$}
      & \multicolumn{2}{c|}{$J_3 = -8$, $J_4 = 23$}
        & \multicolumn{2}{c}{$J_3 = -4$, $J_4 = 17$}
 \\ \hline
& $E$ [K] & $k$ & $E$ [K] & $k$ & $E$ [K] & $k$ & $E$ [K] & $k$ & $E$ [K] & $k$
\\ \hline
$S=8$ & $56.52$  & $\pi$     & $58.07$  & $\pi$     & $57.13$  & $\pi$     & $68.12$  & $\pi$     & $78.39$  & $\pi$ \\
      & $59.49$  & $\pi$     & $61.08$  & $\pi$     & $60.18$  & $\pi$     & $69.81$  & $0$       & $82.26$  & $0$ \\
      & $61.07$  & $0$       & $61.78$  & $0$       & $62.37$  & $0$       & $73.55$  & $\pi$     & $83.49$  & $\pi$
\\ \hline
$S=9$ & $28.48$  & $\pm\pi/2$& $29.15$  & $\pm\pi/2$& $28.92$  & $\pm\pi/2$& $33.99$  & $\pm\pi/2$& $39.13$  & $\pm\pi/2$ \\
      & $44.47$  & $\pi$     & $40.71$  & $\pi$     & $48.22$  & $\pi$     & $35.76$  & $\pi$     & $45.43$  & $\pi$  \\
      & $91.46$  & $0$       & $81.82$  & $0$       & $102.96$ & $0$       & $65.11$  & $0$       & $77.12$  & $0$ \\
      & $119.67$ & $\pm\pi/2$& $113.42$ & $\pm\pi/2$& $132.51$ & $\pm\pi/2$& $174.55$ & $\pm\pi/2$& $124.63$ & $\pm\pi/2$ \\
      & $159.61$ & $\pi$     & $154.43$ & $\pi$     & $175.62$ & $\pi$     & $267.13$ & $\pi$     & $179.93$ & $\pi$ \\
      & $304.45$ & $0$       & $297.02$ & $0$       & $308.05$ & $\pi$     & $501.03$ & $0$       & $436.52$ & $0$
\\ \hline
$S=10$
      & $293.74$ & $\pm\pi/2$& $285.63$ & $\pm\pi/2$& $295.02$ & $\pm\pi/2$& $507.87$ & $\pi$     & $435.95$ & $\pi$ \\
      & $297.30$ & $\pi$     & $289.27$ & $\pi$     & $297.79$ & $\pi$     & $509.01$ & $\pm\pi/2$& $436.83$ & $\pm\pi/2$
\\ \hline
$S=11$& $285.58$ & $0$       & $274.17$ & $0$       & $287.41$ & $0$       & $510.38$ & $0$       & $429.66$ & $0$ \\
      & $303.23$ & $\pm\pi/2$& $313.84$ & $\pm\pi/2$& $290.80$ & $\pm\pi/2$& $696.91$ & $\pm\pi/2$& $537.96$ & $\pm\pi/2$
\\ \hline
$g_{\rm eff}$ &
  \multicolumn{2}{c|}{$1.935$} &
   \multicolumn{2}{c|}{$1.935$} &
    \multicolumn{2}{c|}{$1.92$} &
     \multicolumn{2}{c|}{$2.12$} & 
      \multicolumn{2}{c}{$2.1$} 
\\ \hline
\end{tabular}
\end{table*}

\Section{Magnetic excitations}

\begin{figure}[ht]
\centering
\includegraphics[width=\columnwidth]{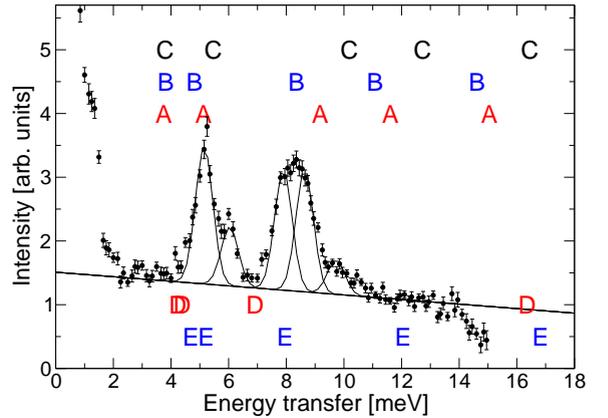}
\caption{Magnetic excitations. Points show the INS spectrum obtained
on MARI with $T = 8$~K, $E_{i} = 17$~meV and $1 \leq Q \leq 2$~{\AA}$^{-1}$.
Lines are Gaussian fits on a linear background. Letters represent
numerical results for the $S = 9$ excitations listed in
Table \ref{TabSpec} for the corresponding parameter choices.
A constant energy $1.29$~meV has been added to all calculated energies
to account for magnetic anisotropy effects \cite{we04}.}
\label{fig:ncomp}
\end{figure}

A number of magnetic excitations in the range of 5 to 35~meV has been
observed by INS experiments performed on different spectrometers
\cite{Hennion97,we04}. The points in Fig.\ \ref{fig:ncomp} show the
spectrum obtained on the MARI spectrometer at ISIS with an incident
energy $E_i = 17$~meV. Five magnetic excitations can be identified
unambiguously in this data, and are shown by the lines in
Fig.\ \ref{fig:ncomp}, which are fits with Gaussian curves on a linear
background. Analysis of their $Q$- and $T$-dependence identifies these
five excitations as magnetic and at least the lowest two of
spin $S = 9$ \cite{we04}. A further magnetic excitation at 27~meV
is the first candidate for an $S = 11$ excitation \cite{we04},
in accord with high-field magnetisation measurements; this energy sets
a lower bound for the numerical calculations

We have performed exact diagonalisation for the model Hamiltonian
(\ref{eqMod}), both to verify the $S = 10$ \GS\ and to determine the
low-lying excitations. The lowest excited states in the sectors with
spin $8 \le S \le 11$ are listed in Table \ref{TabSpec} in ascending
order of energy (for sets A, D and E these extend 
results presented in \cite{we04,Regnault02,Park04}). Spatial symmetry
is described by a momentum $k$ such that the wavefunction acquires a
phase factor ${\rm e}^{ik}$ under a $90^{\circ}$ rotation of the model
in Fig.\ \ref{fig_model}.  The letters in Fig.\ \ref{fig:ncomp} show
the energies of the lowest $S = 9$ excitations for the corresponding data
sets in Table \ref{TabSpec}. A constant shift of $1.29$~meV is added to
all computed energies \cite{we04} to account for the uniaxial cluster
anistropy. Parameter set D \cite{Regnault02} provides only 4 instead of
the observed 5 levels in the energy range of the figure. Although set
E \cite{Park04} yields qualitative agreement with the lowest $S = 9$
excitations, it is not only inconsistent with $\chi$, but also fails to
explain the magnetic excitations observed by INS around 30~meV \cite{we04}.

Parameter sets A, B and C all provide qualitative agreement with the five
$S = 9$ excitations, and the lowest $S = 11$ excitation, observed by INS.
Set A yields the best quantitative agreement \cite{we04}, while B and C
yield an estimate of the error bars: although the individual $J_i$ values
differ by no more than $3.2$~K, the low-lying $S = 9$ levels may shift by
as much as 16~K (1.4~meV) from set A to set C. This demonstrates the high
sensitivity of the excitation spectrum to small changes in the exchange
constants.

Transitions from the $S=10$ \GS\ to states with $S \le 8$ are not
observable by INS due to selection rules. However, our results predict
further low-lying excitations with $S < 9$. In particular, the $S = 8$
excitations in Table \ref{TabSpec} may be interpreted as scattering
states of a pair of the lowest $S = 9$ states ($k = \pm \pi/2$).

\Section{Summary}

In summary, we have determined the microscopic exchange parameters of
Mn$_{12}$-acetate as $J_1 \approx J_2 \approx 65$~K, and $J_3$, $J_4
\approx 5$--10~K. Earlier proposals \cite{Raghu01,Regnault02,Park04} are
inconsistent with the magnetic susceptibility, and do not match our
new INS results. Further improvements to the optimal parameter set
would require a treatment of the uniaxial anisotropies at the
single-ion level, which would be expected to reduce the spread of
the $S = 9$ levels, thereby improving agreement with the INS data.


{\it Acknowledgments:}
This work was supported by the Swiss National Science Foundation, the TMR
programme Molnanomag of the European Union (No: HPRN-CT-1999-00012), and by
the Deutsche Forschungsgemeinschaft through grant SU 229/6-1.





\begin{thebibliography}{00}

\bibitem{Schnack} J.\ Schnack, Lect.\ Notes Phys.\ {\bf 645}, 155 
            (2004).
\bibitem{TuBa02} I.\ Tupitsyn and B.\ Barbara, in {\it Magnetism:
            Molecules to Materials III}, edited by J.S.\ Miller and
            M.\ Drillon, (Wiley-VCH, Weinheim, 2002).
\bibitem{Sessoli93} R.\ Sessoli\etal{, H.-L.\ Tsai, A.R.\ Schake, S.\ Wang,
            J.B.\ Vincent, K.\ Folting, D.\ Gatteschi, G.\ Christou, and
            D.N.\ Hendrickson}, J.\ Am.\ Chem.\ Soc.\ {\bf 115}, 1804 (1993).
\bibitem{Mirebeau99} I.\ Mirebeau\etal{, M.\ Hennion, H.\ Casalta, H.\ Andres,
            H.-U.\ G\"udel, A.V.\ Irodova, and A.\ Caneschi},
            Phys.\ Rev.\ Lett.\ {\bf 83}, 628 (1999).
\bibitem{Raghu01} C.\ Raghu\etal{, I.\ Rudra, D.\ Sen, and S.\ Ramasesha},
            Phys.\ Rev.\ B {\bf 64}, 064419 (2001).
\bibitem{we04} G.\ Chaboussant\etal{, A.\ Sieber, S.\ Ochsenbein,
            H.-U.\ G\"udel, M.\ Murrie, A.\ Honecker, N.\ Fukushima,
            and B.\ Normand}, preprint cond-mat/0404194, to appear in
            Phys.\ Rev.\ B.
\bibitem{Boukhvalov02} D.W.\ Boukhvalov\etal{, A.I.\ Lichtenstein, V.V.\
            Dobrovitski, M.I.\ Katsnelson, B.N.\ Harmon, V.V.\ Mazurenko,
            and V.I.\ Anisimov}, Phys.\ Rev.\ B {\bf 65}, 184435 (2002).
\bibitem{Schake94} A.R. Schake\etal{, H.-L. Tsai, R.J. Webb, K. Folting, G.
            Christou, and D.N. Hendrickson}, Inorg. Chem. {\bf 33}, 6020 (1994).
\bibitem{Regnault02} N. Regnault\etal{, T. Jolicoeur, R. Sessoli, D. Gatteschi,
            and M. Verdaguer}, Phys. Rev. B {\bf 66}, 054409 (2002).
\bibitem{Park04} K. Park\etal{, M.R. Pederson, and C.S. Hellberg},
            Phys. Rev. B {\bf 69}, 014416 (2004).
\bibitem{Barra97} A.L.\ Barra\etal{, D.\ Gatteschi, and R.\ Sessoli}, Phys.\
            Rev.\ B {\bf 56}, 8192 (1997).
\bibitem{Hennion97} M.\ Hennion\etal{, L.\ Pardi, I.\ Mirebeau, E.\ Suard,
            R.\ Sessoli, and A.\ Caneschi}, Phys.\ Rev.\ B {\bf 56},
            8819 (1997).




\end{thebibliography}
\end{document}